\documentstyle[multicol,aps]{revtex}

\begin{document}

\draft

\newcommand{\beq}{\begin{equation}}
\newcommand{\eeq}{\end{equation}}
\newcommand{\bdis}{\begin{displaymath}}
\newcommand{\edis}{\end{displaymath}}
\newcommand{\bea}{\begin{eqnarray}}
\newcommand{\eea}{\end{eqnarray}}
\newcommand{\barr}{\begin{array}}
\newcommand{\earr}{\end{array}}

\title{Dynamics of Eulerian walkers.}

\author{A.M. Povolotsky, V.B. Priezzhev, R.R. Shcherbakov}

\address{
Bogoliubov Laboratory of Theoretical Physics, \\
Joint Institute for Nuclear Research, Dubna 141980, Russia.}

\date{\today}

\maketitle

\begin{abstract}
We investigate the dynamics of Eulerian walkers as a
model of self-organized criticality.
The evolution of the system is subdivided into
characteristic periods which can be seen as avalanches.
The structure of avalanches is described
and the critical exponent in the
distribution of first avalanches $\tau=2$ is determined.
We also study a mean square displacement
of Eulerian walkers
and obtain a simple diffusion law in the critical state.
The evolution of underlying medium
from a random state to the critical one
is also described.
\end{abstract}

\begin{multicols}{2}

\section{Introduction.}
To illustrate the phenomenon of the self-organized
criticality (SOC)\cite{btw}
a wide range of cellular automata such as sand piles, rice piles
and forest fires,
have been proposed \cite{btw,f,ds}.
They assume a system consisting of a large amount of
elements.  The energy
being randomly added to the system is redistributed
among the degrees of freedom by a kind of nonlinear
diffusion.
This is realized by means of avalanche-like processes
which carry the added energy out
of the system.
As a rule, the system spontaneously evolves
towards the critical state free of characteristic
length and time scale.
In this state, probabilistic distributions of quantities,
characterizing the statistical ensemble, exhibit the power
law behavior.

What features of the SOC dynamics are responsible for
the existance of a dynamical attractor in complex systems?
What are the origins of the scaling and self-similarity
in the stationary state that appeares?
To answer these questions, one evidently needs
investigating nonlinear diffusion
in the SOC models and studying the structure of
avalanches. Due to a rather complicated
dynamics of most SOC models, the description
of their evolution is a very difficult problem.
Up to now, the most analytically tractable
model has been the Abelian sandpile model (ASM) \cite{d}.
Due to its simple algebraic structure, the detailed
description of the SOC state of ASM has been given, and
some critical exponents have been found \cite{md,p,ikp,pki}.

Recently, a new model has been proposed
which is called the Eulerian walkers model (EWM)\cite{pddk}.
It also possesses
Abelian properties and exhibits SOC.
While EWM has many common features with
ASM, its dynamics is much more transparent.
This model is formulated as a deterministic walk
interacting with a changeable medium.
The motion of a particle is affected by the medium,
and in its turn affects the medium inducing strong
correlations in the system.
If the walk occurs in a closed system, it continues
infinitely long and eventually gets
self-organized into Eulerian trails.
If a system is open, the particles can leave the
system and new particles drop time after time.
In this case, the system evolves to the critical
state similar to that in ASM.
By analogy with ASM, the avalanches in EWM
have been introduced \cite{pprep} as periods
of reconstruction of recurrent states,
after they have broken by an added particle.

In this article, we use the properties of Eulerian trails
to describe the
evolution of the system in detail.
In this way, we find the critical exponent
in the distribution of first avalanches.
Using the Green function of EWM introduced in \cite{pddk},
we obtain the critical exponent characterizing
spread of particles with time.

\section{Algebraic properties of Eulerian walkers model.}

The Eulerian walkers model is defined as follows.
Consider an arbitrary connected graph
$\bf G$ consisting of $N$ sites.
Each site of $\bf G$ is associated with
an arrow which is directed along one of the incident bonds.
The arrow directions at the site $i$
are specified by the integers $n_i$, ($ 1 \leq n_i \leq \tau_i $) where
$\tau_i$ is the number of nearest neighbors of the site $i$.
The set \{$n_i$\} gives a complete description of the medium.
Starting with an arbitrary arrow configuration
one drops the particle to a site of $\bf G$ chosen at random.
At each time step:\\
$(i)$ the particle arriving at a site $i$
changes the arrow direction from $n_i$ to $n_i+1(mod\tau_i)$\\
$(ii)$ the particle moves one step along the new arrow direction
from $i$ to the neighboring site $i'$.

Having no endpoints on
$\bf G$, the particle continues to walk
infinitely long. Due to a finite number of possible states
of the system, it eventually
settles into the Poincare cycle.
For most dynamic systems
the recurrence time of this cycle grows exponentially
with $N$. It has been shown in \cite{pddk}
that for the Eulerian walk the Poincare cycle is squeezed
to the Eulerian trail  \cite{hp} with
the recurrence time of an order of $N$.
During the Eulerian trail
the particle passes all bonds of the graph
exactly once in each direction.

Let $\bf G$ be an open graph.
It means that one auxiliary site is introduced which is
called a sink.
The subset of sites of $\bf G$ connected with the
sink forms an open boundary.
The sink does not have an arrow and
the particle reaching the sink leaves the system.
After that, a new particle is dropped to
the a site of $\bf G$ chosen at random.
Since on the closed graph the particle
visits all sites during the walk, at the open one
it ever reaches the sink.
A set \{$C$\} of configurations $C=\{n_j\}$ which
remains on $\bf G$ after the particle left $\bf G$
for the sink, is the set of stable configurations.
The operator $a_i$ can be introduced
\beq
a_iC=C'
\eeq
which describes the resulting transformation
caused by dropping the particle to the site $i$.
As usual in the theory of Markov
chains, the set \{$C$\} may be divided into two subsets.
The first subset denoted by $\{R\}$ includes those
configurations which can be obtained from an arbitrary
configuration by a sequential action by the operators $a_{i}$. It
follows from the definition that the subset $\{R\}$ is closed
under a multiple action by the operators $a_{i}$. Once the system gets
into $\{R\}$, it never gets out under subsequent evolution.  All
nonrecurrent configurations are called transient and form the
subset $\{T\}$ which is the complement to the set $\{R\}$.
By definition any recurrent configuration $C \in \{R\}$ may be reached
from any another $C' \in \{R\}$ by a subsequent action of the
operators $a_i$.
Since this is valid for $C'= C$ too,
the identity operator acting in  $\{R\}$ exists.
In addition,
the operators $a_i$ have the following properties:

{\bf 1}. For arbitrary sites $i$ and $j$
 and for any configuration of arrows $C$
\begin{equation}
\label{2}
a_{i}a_{j} C = a_{j}a_{i} C
\end{equation}

{\bf 2}. For any recurrent configuration $C
 \in \{ R \}$, there exists a unique $$ (a_{i}^{-1} C) \in \{ R
\}$$ such that
\begin{equation}
\label{3}
a_{i}( a_{i}^{-1} ) C =  C
\end{equation}

\noindent
The proof of these statements is similar to ones
for the avalanche operators in ASM \cite{d}
and is given in \cite{p}.
Thus, the operators  $a_i$ acting in the set of
recurrent configurations
$\{R\}$ form the Abelian group.
The addition of $\tau_i$ particles to site $i$
gives the same effect as the addition of one particle to each
of $\tau_i$ neighbors of $i$.
It returns the arrow outgoing from $i$ to the former position
and initiates the motion of one particle to each neighboring
site. In the operator form this looks like
\beq
{a_i}^{\tau_i}=\prod_{k=1}^{\tau_i}a_{j_k}
\label{rel1}
\eeq
where $j_k$ are the neighbors of the site $i$.
Introducing the discrete Laplacian
on $\bf G$ as
\beq
\Delta_{ij} = \left\{
\begin{array}{ccc}
 \tau_i & , & i=j \\
{-1} & , & i
{\rm ~and~} j {\rm ~are~connected~by~bond} \\
0 & , & {\rm otherwise}.
\end{array}
\right. \label{Laplacian}
\eeq
and using Eq.(\ref{rel1}), one can write
the identity operator as
\beq
E_i=\prod_{j \in {\bf G}}a_j^{\Delta_{ij}}.
\label{unit}
\eeq
Since all recurrent configurations can be obtained
from an arbitrary one
by a successive action by operators $a_i$, one can represent
any $C\in\{R\}$ in the form
\beq
C=\prod_{i \in {\bf G}}(a_i)^{n_i}C^*.
\eeq
The $N$-dimensional vector $\bf n$ labels all possible recurrent
configurations.
Eq. (\ref{unit}) shows that two vectors $\bf n$ and $\bf n'$
label the same configuration if the difference between
them is $\sum_jm_j\Delta_{ij}$ where $m_j$ are integers.
The $N$-dimensional space $\{{\bf n}\}$ has a periodic structure
with an elementary cell of the form of a hyper-parallepiped
with base edges $\vec{e}_i=(\Delta_{i1}, \Delta_{i2},...,\Delta_{iN})$.
Thus, the number of non-equivalent recurrent configurations is
\beq
N=det\Delta,
\eeq
which is Kirchhoff's formula \cite{k} for spanning trees
and Dhar's
formula for ASM~\cite{d}.
The correspondence to ASM is not surprising. The
algebra of the operators $a_{i}$ completely coincides with that
of avalanche operators of the Abelian sandpile model
\cite{d}. Moreover, the identity operator (\ref{unit}) has the
same form for both the models . This is the reason why the
numbers of recurrent configurations coincide.

Continuing the analogy between EWM and
sandpiles, one can find the expected number $G_{ij}$ of full
rotations of the arrow at site $j$, due to the particle dropped
at $i$ \cite{d}. During the walk, the expected number of
steps from $j$ is $\Delta_{jj} G_{ij}$ whereas $-\sum_{k
\not=j}G_{ik}\Delta_{kj}$ is the average flux into $j$. Equating
both the fluxes, one gets
\begin{equation}
\label{15}
\sum\limits_{k}G_{ik} \Delta_{kj} = \delta_{ij}
\end{equation}
or
\begin{equation}
\label{16}
G_{ij} = [ \Delta^{-1} ]_{ij}.
\end{equation}
The expected number of full rotations of the arrow
is equal to the number of entries into the site $j$
divided by $\tau_j$.
On the other hand, the number of visits
of a site for the random walk is also
Green function of the Laplace equation.
Thus, we have a surprising fact
that the number of visits of the site
by the particle for the deterministic motion in EWM
coincides with that for the  ordinary random walk.

The direct correspondence between spanning trees, recurrent
configurations of EWM and Eulerian trails can be established
in the following way.
The walking particle leaves each
site along an arrow after turning the arrow.
Therefore,
the trajectory of the particle is a
trace of arrows.
If $\bf G$ is an open graph,
all trajectories end at the sink and never form loops.
The corresponding arrow configurations are acyclic ones.

Given an acyclic arrow configuration,
we can construct a unique spanning tree rooted in
the sink and vice versa.
Indeed, all bonds along which the arrows are directed
form the spanning tree. Conversely,
if we have a spanning tree rooted in the sink we can obtain the
acyclic arrow configuration by pointing the arrow from each site
along the path leading to the sink.
This correspondence allows us to identify
the acyclic arrow configurations and spanning trees.
Further, we do not distinguish between them.
Saying spanning tree we mean both the spanning
tree and its arrow representation.

If $\bf G$ is the closed graph, the particle settles into
the Eulerian trail during which it
passes each bond exactly once in each direction.
Let the particle, which has already visited all sites,
arrive at a site $i$ at some moment.
If we now remove the arrow from $i$,
we obtain the acyclic arrow configuration where
arrow paths from any site end in $i$.
This defines the spanning tree rooted in the site
of the current particle location.
Therefore, given a site $i$, each Eulerian trail on a
closed graph has the unique
spanning tree representation.

\section{Avalanche dynamics.}

The particle added to the recurrent configuration of
ASM may induce successive topplings of sites
called the avalanche.
At the initial moment, it destroys the recurrent configuration
and the system leaves the critical state.
After the avalanche stops, the recurrent configuration is
restored again.
Thus, the avalanche in ASM may be defined as a period
of reconstruction of the recurrent state.
This definition may be directly applied to EWM.

We start with a recurrent state of EWM.
The corresponding arrow configuration
forms a spanning tree.
Once a particle is dropped, it may destroy the
spanning tree by closing a loop of arrows.
During the evolution, one loop can be transformed into another.
When all loops disappear, the spanning tree is restored.
The interval of existance of the loop can be called
the {\it avalanche of cyclicity} or simply avalanche.
The loops may be created and destroyed several times
during the motion of one particle.
Therefore, unlike ASM, an addition of one particle
may initiate several avalanches in the system.
When a particle comes to the sink, it always directs the
arrow to the sink thus restoring the spanning tree. Therefore, when the
particle leaves the system the avalanche always ends and recurrent
state is restored. All motions of the particle represent successive
transitions from one recurrent state to another through
avalanches. To study the evolution of the system,
the structure of the avalanche should be considered in detail.

Consider
the Eulerian walk on the square lattice $\cal L$
of size $ L \times L $ with open boundary conditions.
Each boundary site is connected to the sink by one bond
on the edge and by two bonds at the corners of $\cal L$.
The rule of arrow rotations is the same for all sites.
If we denote the bonds outgoing from a site $i$ by
$N,E,S,W$, the rule of rotations is
$N \rightarrow E \rightarrow S \rightarrow W$.
In other words,
when the particle arrives at a site, the arrow outgoing
from this site turns to the next bond clockwise.
Due to a topological reason,
this rule leads to a simple
structure of avalanches, namely to
compactness of clusters of sites visited by the particle.

Let the particle be dropped to a recurrent configuration which
is a spanning tree. At some step the first loop is created.
The arrows can form loops of two kinds: clockwise and anti-clockwise.
The loop is clockwise if tracing the loop along arrows leaves
the interior of the loop on the right and
anti-clockwise otherwise.
It is easy to see that due to the clockwise rule of rotations,
only clockwise loops may be created
from recurrent states.
Indeed, the anti-clockwise loop arises when
the arrow, which closes this loop,
is directed at the previous time step into the area bounded
by the loop. The arrow path beginning
from this arrow could not leave the area of the loop
without intersections with the loop.
This means that before this loop was closed,
another loop existed,  which contradicts the assumption
that we start with a spanning tree.

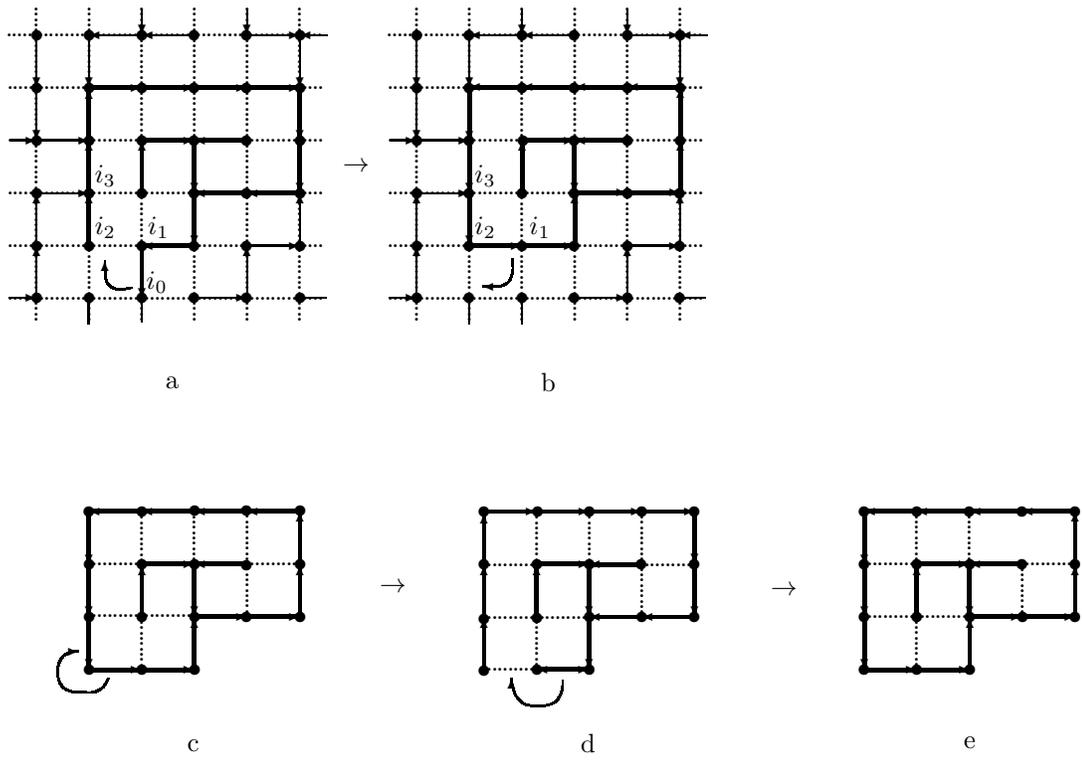
\begin{figure}
\caption{(a) -- closing the loop at $i_1$.
(b) -- the last step before openning the loop.
(c), (d), (e) - the evolution on the closed graph settled into the
Eulerian trail.
The loops in (a) and (b) exactly coincide with those in (d) and (e).}
\label{aval}
\end{figure}

Consider the evolution after
the closing of a clockwise loop at the spanning tree.
Denote by
$ij$ the arrow
if it is pointed from site $i$
to site $j$.
Analogously, we denote by $i_1i_2i_3...$ the arrow path
if the arrow from site $i_1$ is pointed to site $i_2$, the arrow
from $i_2$ is pointed to $i_3$ and so on.
Let a spanning tree exist at the time step $(t-1)$,
while at the step $t$, the particle arrived at the site
$i_1$ changes the arrow direction
from $i_1i_0$ to $i_1i_2$ and the clockwise loop
${\cal O}^+ = i_1i_2i_3...i_ni_1$
appears (Fig~\ref{aval}a).
Now, we can prove the following:\\
{\bf Proposition 1:}
The particle does not leave the area of the loop ${\cal O}^+$
and the spanning tree cannot be restored
until all arrows inside the loop area make the full rotation
and the arrows belonging the loop itself change the direction
to anti-clockwise forming the anti-clockwise loop
${\cal O}^- = i_1i_n...i_2i_1$.
At the last step when $\cal{O}^-$ appears,
the particle arrives at $i_2$ and at the
next step the arrow at $i_2$ rotates out of the loop area
and the spanning tree may be restored
(Fig.~\ref{aval}b).

Proof:
Consider EWM
on the auxiliary graph $\cal G$,
which is a part of the square lattice bounded by the loop
${\cal O}^-$ with closed boundary conditions.
The closed boundary
means that all bonds that link boundary sites
$i_1, i_2, i_3,..., i_n$ with the sites of the lattice
outside the loop area are removed.
The rules of rotations are modified in such a way
that an arrow skips deleted bonds.
We consider the Eulerian trail at $\cal G$
starting from the site $i_2$. At the initial
moment, the arrow
configuration at $\cal G$
differs
from that on the lattice $\cal L$ only by orientation of the loop:
instead of the clockwise loop ${\cal O}^+ = i_1i_2i_3...i_ni_1$
on $\cal L$, we have the anti-clockwise loop
${\cal O}^- = i_1i_n...i_2i_1$ on $\cal G$  (fig.\ref{aval}c).
Starting from the first step,
$(n-1)$ successive steps reverse ${\cal O}^-$ into ${\cal O}^+$
and the particle arrives at $i_1$  (fig.\ref{aval}d).
Notice that the initial arrow configuration on $\cal G$
corresponds to that described in the previous
section, when the particle has already settled into the Eulerian trail
on the closed graph.
Indeed, at the first moment,
all arrows except the arrow at the current particle
location site form the spanning tree rooted in this site.
Hence,
the subsequent
evolution leads again to the loop ${\cal O}^-$
via full rotation of arrows at all internal sites (fig.\ref{aval}e).
On the other hand,
this part of evolution of the graph $\cal G$ coincides
with one on the original lattice $\cal L$
since the moment when the loop ${\cal O}^+$ is closed (Fig.~\ref{aval}a)
up to the moment when it is changed by ${\cal O}^-$  (Fig.~\ref{aval}b).
At the last step $i_2i_1$ rotates out of the loop area
and the loop may be broken.
Before this moment the loop exists permanently
as during Eulerian trail one loop always exists.
The proposition is proved.

Generally, the avalanche does not
necessarily end
after that.
Two situations are possible.
At the last step, the arrow at $i_2$ turns outside the
anti-clockwise loop
$i_2i_1 \rightarrow i_2i_2'$.
If $i_1'$ is connected to the sink through
the arrow path, the spanning tree is restored
and the avalanche is finished.
This is the case of a one-loop avalanche.
In the other case, the arrow path from $i_2'$ goes
to $i_2$, $i. e.$ $i_2'$  is the predecessor of
$i_2$ with respect to the sink.
Then, one more loop is closed and the avalanche continues.
This is a two-loop avalanche.
The second loop relaxes
like the first one.
When the second loop is reversed, the
spanning tree is always restored
because at the last step the particle arrives at $i_0$
which was connected to the sink by an arrow path
before the avalanche started.

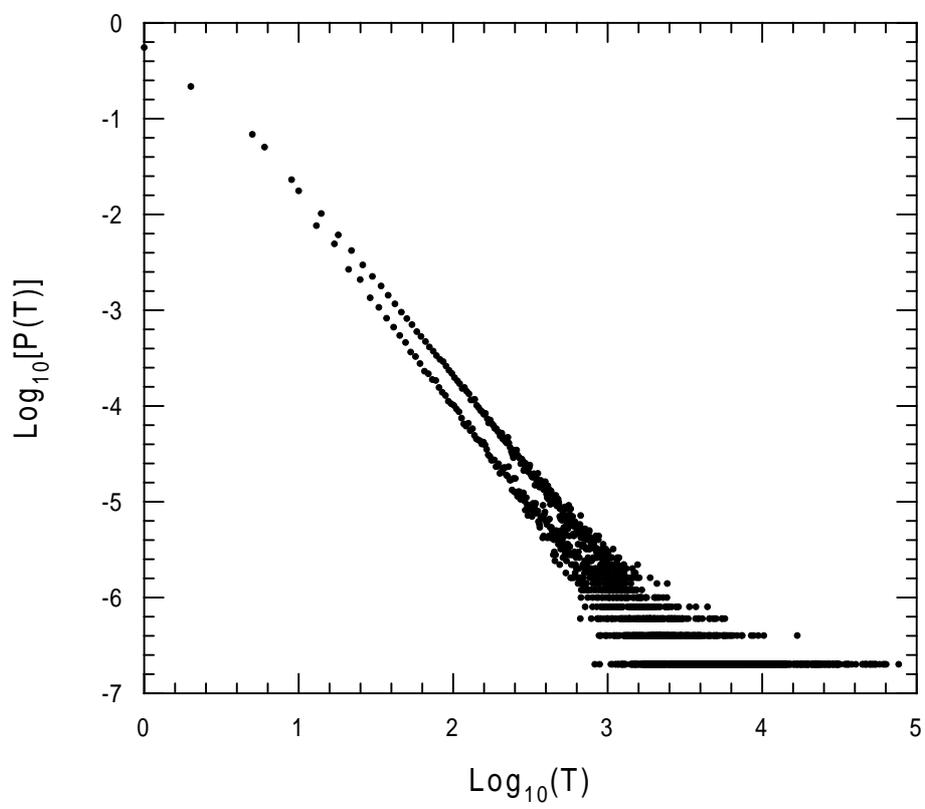
\begin{figure}
\caption{The distribution of duration of the first avalanche in the
SOC state is shown on the double logarithmic plot.
The distribution splits into two parts as described in the text.
The slope of both the
parts is the same with the critical exponent $\tau=2.0$.}
\label{firstav}
\end{figure}

Several consequences may be obtained
from the picture described.
During the avalanche the particle visits
sites inside the loop four times,
sites of the edge two times,
one time at the corner $\frac{\pi}{2}$ and three times
at the corner $\frac{3\pi}{2}$.
Then, the duration of relaxation of a loop
is given by the formula
\beq
T=(4s+2p-4)+1
\eeq
where $s$ is the number of inner sites, and
$p$ is the perimeter of the loop.
As the avalanches may consist of one or two loops,
the duration of avalanches may be equal to
\bea
T_1=2k+1  \nonumber  \\
T_2=2k+2
\eea
where $k=0,1,2,...$
This explains the double distribution
of durations of avalanches (Fig.~\ref{firstav}) obtained in~\cite{spp}.
Also we can find the critical exponent of the duration
distribution for the first avalanche.
In the thermodynamic limit, the duration of avalanches
grows as the area of the loop.
It has been shown in~\cite{mdm} that the probability
to get a loop of the size $s$
when a bond is added to the spanning tree
at random is equal to
\beq
P(s) \sim s^{-\frac{11}{8}}.
\label{dhar}
\eeq
The distinction of our case from this one
is that the loop is closed by turning the single arrow
that was connected to the sink
through an arrow path before the turn.
Hence, the distribution (\ref{dhar}) should be divided
by the perimeter of the loop.
Taking into account that that perimeter scales with the linear
size as a
fractal
dimension of a chemical path $p \sim r^\frac{5}{4}$ and
that the loop is compact $s \sim r^2$,
we obtain
\beq
{\cal P}(s) \sim \frac{s^{-\frac{11}{8}}}{r^\frac{5}{4}}\sim s^{-2}.
\eeq
Thus, for the first avalanches
the critical exponent of the distribution of duration is
$\tau=2$.
The one- and two-loop avalanches differ
only in a local structure of the spanning tree
at the site of closing the loop.
Therefore, the critical exponents are the same for both the distributions.
This result is in excellent agreement with numerical simulations
presented in Fig.~\ref{firstav} where we have considered the EWM
on the square lattice of linear size $L=400$ with open boundary
conditions.

The evaluation of the exponent of the first avalanche size
distribution in EWM is similar to one for the
distribution of sizes of erased loops in the loop-erased
walks, which was studied in \cite{dd}.
The same exponent $\tau=2$ was obtained.

The result
$\tau=2$ is valid only for the first avalanches for their
independence of each other.
The analytical derivation
of $\tau$ for arbitrary avalanches
is a more difficult problem due to
correlations between subsequent avalanches
appearing during the evolution of one particle.

\section{Propagation of Eulerian walkers.}

Besides the evolution of the system as a whole,
we can describe the
 motion of
the particle itself.
Consider the particle dropped on the lattice with a spanning tree.
We call the site $i$ a predecessor of $j$
if the arrow path comes from $i$ to $j$.
Since the particle motion is traced by a path of arrows,
all visited sites are predecessors of the site of a current
particle location.
If the particle arrives at the site which is its predecessor,
the loop is closed.
Thus, the particle can visit the sites that have already been visited
only during an avalanche.

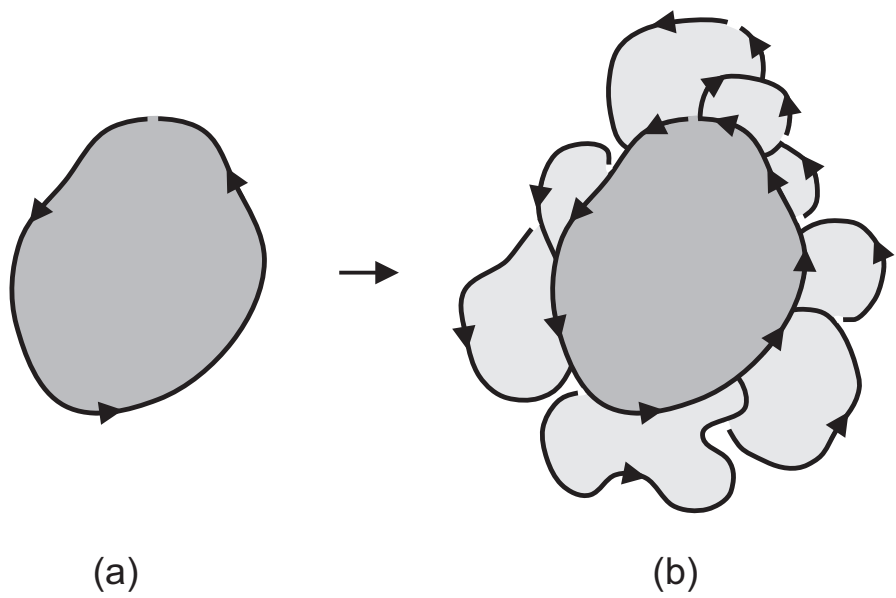
\begin{figure}
\caption{A subsequent evolution of a cluster
of visited sites in the SOC state.
A schematic picture of the cluster after the first (a) and second
(b) stages of evolution. The areas with different
numbers of visits are shown by different colors. The directions of arrows
correspond to their final positions.}
\label{rose}
\end{figure}

We subdivide the motion of the particle into the following
stages. The first stage coincides with the first avalanche.
At the moment it finishes,
the avalanche area remains bounded by the anti-clockwise loop
opened at the bond connecting two sites where it
begins and where it ends.
Further, moving on the lattice, the particle cannot enter
the area of the first avalanche.
New avalanches appear
beyond the first one being attached to its boundary and
tending to go clockwise around it.
Eventually, the particle creates a loop enclosing the area
of the first avalanche. When the avalanche corresponding to this loop ends,
the second stage of the evolution finishes.
At this moment, we have the cluster of visited sites
which consists of the area of the
first avalanche, where each inner site is visited eight times,
surrounded by the clusters of subsequent avalanches, where all
sites are visited four times (Fig.~\ref{rose}).

The further behavior of the system is similar.
If at some evolution stage we have a cluster of visited sites,
at the next stage all sites of this cluster
will be visited four more times and
some new area will be added to the cluster of visited
sites.
After each evolution stage finishes,
the cluster of visited sites
is compact because it consists of compactly
situated avalanche clusters.

Thus, we obtain the system of compact clusters
where the sites are visited $4N,\,\,\, N=1,2,...$ times.
The clusters are strictly embedded one into another with a
growing number of visits like
Grassberger-Manna clusters in ASM \cite{gm}.

Using this picture, we can find time dependence
of the mean square displacement of the particle
in the critical state.
The number of visits ($N(R)$)
of a site separated from the origin by the distance $R$
is given by the Green function
of the Laplace equation Eq.(\ref{15}).
When $|{\bf r-r'}|$ tends to the lattice size, $G({\bf r,r'})$
decays as $\log(L/|{\bf r-r'}|)$, so we can write
\beq
\frac{d N(R)}{d R} \sim -\frac{1}{R}.
\label{increase}
\eeq

On the other hand, the time $T$ required for the
particle  to
visit four times all the sites of the
compact cluster, is of an order of its size $R^2$.
Then, the velocity of the growth is
\beq
\frac{d N}{d T} \sim -\frac{1}{R^2}
\label{time}
\eeq
Using (\ref{increase}) and (\ref{time}) and the property of
the embedded clusters, we obtain
the mean square displacement
\beq
<R^2> \sim T^{2\nu}, \nu=\frac{1}{2},
\label{dif}
\eeq
that is the diffusion law of the simple random walk.

In the transient state, we have no the spanning tree
representing the evolution of the system.
The sites already visited by the particle are
connected with the current particle location
by an arrow path and the cluster of these sites has an
acyclic structure.
The cluster of acyclic arrows is embedded into the media
of randomly distributed arrows.
When the particle enters the cluster
of visited sites it behaves like in the critical state.
Each time the linear size of the visited cluster increases by $\Delta$,
the particle visits all sites of the cluster four times again.
The time of increasing $\Delta$ is of order of the size of the cluster,
$i.e.$
\beq
\frac{d T}{d R} \sim R^2.
\eeq
Thus, instead of the simple diffusion law (\ref{dif}) in the critical state,
for the transient states we obtain
\beq
<R^2> \sim T^{2\nu_t}, \nu_t=\frac{1}{3}
\label{random}
\eeq
which has already been determined in \cite{pddk}.
Note that the power law (\ref{random})
is valid only at the time scale much greater than the
time spent inside one cluster of the visited sites.
Inside the cluster, the motion of the particles is similar
to that in the critical state with the diffusion law (\ref{dif}).
By averaging over a large number of returns
to the origin one obtains $\nu_t=\frac{1}{3}$.

Now we can estimate the average
time required to reach the critical state starting from an
arbitrary random configuration of arrows.
In order to get a spanning tree on the lattice, the particle
must visit all sites at least once.
Using (\ref{random})
we can obtain for the
lattice of the size
$L \times L$
\beq
T_c\sim L^3.
\eeq
The same time is required for the particle
walking on the closed graph to settle into the Eulerian trail.

We also measured numerically the motion of the particle
in the system. Starting from the transient state, the mean square
displacement of the particle is described by the power law with
the critical exponent $\nu_t=0.33$ as is shown in Fig.~\ref{displace}a.
The subsequent evolution
of the system by the repetitive addition of particles
changes this power law. For the system in the SOC state,
we obtained the value $\nu = 0.5$ (Fig.~\ref{displace}b).
These simulations illustrate very well the exact results
obtained above.

In summary, we considered the dynamics of the Eulerian Walkers Model.
The structure of avalanches in the SOC state was studied
in detail. We obtained the critical exponent for the distribution
of durations of the first avalanche.
Considering the evolution of the system as a sequence of avalanches,
we found the simple diffusion law for the mean square displacement
of the particle in the SOC state. The crossover from the transient
state into the SOC state was described as well. The obtained exact
results were confirmed by numerical simulations.

\vspace{2cm}

\section*{Acknowledgments}

We would like to thank Prof. D.~Dhar for fruitful discussions.
Partial financial support by the Russian Foundation for Basic Research
under grant No.~97-01-01030 and by the
INTAS under grant No.~96-690 is acknowledged.

\end{multicols}

\newpage

\begin{figure}
\caption{The dependence of the mean square displacement of the particle
on time in the transient (a) and SOC (b) states.
The obtained values of the critical exponents are
$\nu_t=0.33$ and $\nu=0.5$, respectively.}
\label{displace}
\end{figure}

\newpage

\setcounter{figure}{0}

\begin{figure*}
\unitlength=0.35mm
\special{em:linewidth 0.8pt}
\linethickness{0.8pt}
\begin{picture}(469.21,435.88)
\linethickness{0.5pt}
\bezier{84}(100.11,344.00)(89.44,341.33)(90.44,351.00)
\put(90.44,351.00){\vector(0,1){3.33}}
\linethickness{1.4pt}
\put(104.44,360.00){\vector(0,-1){20.00}}
\put(124.44,360.00){\vector(-1,0){20.00}}
\put(124.44,380.00){\vector(0,-1){20.00}}
\put(144.44,380.00){\vector(-1,0){20.00}}
\put(164.44,400.00){\vector(0,-1){20.00}}
\put(164.44,420.00){\vector(0,-1){20.00}}
\put(144.44,420.00){\vector(1,0){20.00}}
\put(124.44,420.00){\vector(1,0){20.00}}
\put(104.44,420.00){\vector(1,0){20.00}}
\put(84.44,420.00){\vector(1,0){20.00}}
\put(84.44,400.00){\vector(0,1){19.67}}
\put(84.44,380.00){\vector(0,1){20.00}}
\put(84.44,360.00){\vector(0,1){20.00}}
\put(164.44,380.00){\vector(-1,0){20.00}}
\put(104.44,380.00){\vector(0,1){20.00}}
\put(124.44,400.00){\vector(0,-1){20.00}}
\put(104.44,400.00){\vector(1,0){20.00}}
\put(144.44,400.00){\vector(-1,0){20.00}}
\
\linethickness{0.4pt}
\
\put(64.44,380.00){\vector(1,0){20.00}}
\put(64.44,440.00){\vector(0,-1){20.00}}
\put(64.44,420.00){\vector(0,-1){20.00}}
\put(64.44,400.00){\vector(1,0){20.00}}
\put(54.44,400.00){\vector(1,0){10.00}}
\put(54.11,340.00){\vector(1,0){10.33}}
\put(64.44,340.00){\vector(0,1){20.00}}
\put(64.44,360.00){\vector(0,1){20.00}}
\put(84.44,440.00){\vector(0,-1){20.00}}
\put(104.44,440.00){\vector(-1,0){20.00}}
\put(104.44,449.67){\vector(0,-1){9.67}}
\put(124.44,440.00){\vector(-1,0){20.00}}
\put(144.44,449.67){\vector(0,-1){9.67}}
\put(144.44,440.00){\vector(1,0){20.00}}
\put(164.44,440.00){\vector(0,-1){20.00}}
\put(174.77,440.00){\vector(-1,0){10.33}}
\put(124.44,340.00){\vector(1,0){20.00}}
\put(144.44,340.00){\vector(0,1){20.00}}
\put(144.44,360.00){\vector(1,0){20.00}}
\put(164.44,360.00){\vector(0,1){20.00}}
\linethickness{0.4pt}
\
\linethickness{1.4pt}
\put(248.88,380.00){\vector(0,1){20.00}}
\put(268.88,400.00){\vector(0,-1){20.00}}
\put(248.88,400.00){\vector(1,0){20.00}}
\put(288.88,400.00){\vector(-1,0){20.00}}
\put(228.65,359.91){\vector(1,0){20.14}}
\put(248.79,359.91){\vector(1,0){20.14}}
\put(268.93,359.91){\vector(0,1){20.14}}
\put(268.93,380.05){\vector(1,0){19.91}}
\put(308.98,380.05){\vector(0,1){19.91}}
\put(308.98,400.20){\vector(0,1){19.91}}
\put(308.98,420.11){\vector(-1,0){20.14}}
\put(288.84,420.11){\vector(-1,0){19.91}}
\put(269.16,420.11){\vector(-1,0){20.37}}
\put(248.79,420.11){\vector(-1,0){19.91}}
\put(228.88,419.87){\vector(0,-1){20.14}}
\put(228.88,399.73){\vector(0,-1){19.68}}
\put(228.88,380.05){\vector(0,-1){20.14}}
\put(288.84,380.05){\vector(1,0){20.14}}
\linethickness{0.4pt}
\put(208.88,380.00){\vector(1,0){20.00}}
\put(208.88,440.00){\vector(0,-1){20.00}}
\put(208.88,420.00){\vector(0,-1){20.00}}
\put(208.88,400.00){\vector(1,0){20.00}}
\put(198.88,400.00){\vector(1,0){10.00}}
\put(198.55,340.00){\vector(1,0){10.33}}
\put(208.88,340.00){\vector(0,1){20.00}}
\put(208.88,360.00){\vector(0,1){20.00}}
\put(228.88,440.00){\vector(0,-1){20.00}}
\put(248.88,440.00){\vector(-1,0){20.00}}
\put(248.88,449.67){\vector(0,-1){9.67}}
\put(268.88,440.00){\vector(-1,0){20.00}}
\put(288.88,449.67){\vector(0,-1){9.67}}
\put(288.88,440.00){\vector(1,0){20.00}}
\put(308.88,440.00){\vector(0,-1){20.00}}
\put(319.21,440.00){\vector(-1,0){10.33}}
\put(268.88,340.00){\vector(1,0){20.00}}
\put(288.88,340.00){\vector(0,1){20.00}}
\put(288.88,360.00){\vector(1,0){20.00}}
\put(308.88,360.00){\vector(0,1){20.00}}
\linethickness{0.5pt}
\bezier{84}(244.87,354.58)(246.56,343.46)(236.78,344.47)
\put(238.13,344.47){\vector(-1,0){4.38}}
\
\linethickness{1.4pt}
\put(294.14,238.76){\vector(-1,0){19.69}}
\put(274.44,198.89){\vector(-1,0){20.00}}
\put(274.44,218.89){\vector(0,-1){20.00}}
\put(294.44,218.89){\vector(-1,0){20.00}}
\put(314.44,238.89){\vector(0,-1){20.00}}
\put(314.44,258.89){\vector(0,-1){20.00}}
\put(294.44,258.89){\vector(1,0){20.00}}
\put(274.44,258.89){\vector(1,0){20.00}}
\put(254.44,258.89){\vector(1,0){20.00}}
\put(234.44,258.89){\vector(1,0){20.00}}
\put(234.44,238.89){\vector(0,1){19.67}}
\put(234.44,218.89){\vector(0,1){20.00}}
\put(234.44,198.89){\vector(0,1){20.00}}
\put(314.44,218.89){\vector(-1,0){20.00}}
\put(254.44,218.89){\vector(0,1){20.00}}
\put(274.44,238.89){\vector(0,-1){20.00}}
\put(254.44,238.89){\vector(1,0){20.00}}
\
\linethickness{1.4pt}
\
\put(398.88,218.89){\vector(0,1){20.00}}
\put(418.88,238.89){\vector(0,-1){20.00}}
\put(398.88,238.89){\vector(1,0){20.00}}
\put(438.88,238.89){\vector(-1,0){20.00}}
\put(378.65,198.80){\vector(1,0){20.14}}
\put(398.79,198.80){\vector(1,0){20.14}}
\put(418.93,198.80){\vector(0,1){20.14}}
\put(418.93,218.94){\vector(1,0){19.91}}
\put(458.98,218.94){\vector(0,1){19.91}}
\put(458.98,239.09){\vector(0,1){19.91}}
\put(458.98,259.00){\vector(-1,0){20.14}}
\put(438.84,259.00){\vector(-1,0){19.91}}
\put(419.16,259.00){\vector(-1,0){20.37}}
\put(398.79,259.00){\vector(-1,0){19.91}}
\put(378.88,258.76){\vector(0,-1){20.14}}
\put(378.88,238.62){\vector(0,-1){19.68}}
\put(378.88,218.94){\vector(0,-1){20.14}}
\put(438.84,218.94){\vector(1,0){20.14}}
\linethickness{1.4pt}
\put(104.44,218.89){\vector(0,1){20.00}}
\put(124.44,238.89){\vector(0,-1){20.00}}
\put(104.44,238.89){\vector(1,0){20.00}}
\put(144.44,238.89){\vector(-1,0){20.00}}
\put(84.21,198.80){\vector(1,0){20.14}}
\put(104.35,198.80){\vector(1,0){20.14}}
\put(124.49,198.80){\vector(0,1){20.14}}
\put(124.49,218.94){\vector(1,0){19.91}}
\put(164.54,218.94){\vector(0,1){19.91}}
\put(164.54,239.09){\vector(0,1){19.91}}
\put(164.54,259.00){\vector(-1,0){20.14}}
\put(144.40,259.00){\vector(-1,0){19.91}}
\put(124.72,259.00){\vector(-1,0){20.37}}
\put(104.35,259.00){\vector(-1,0){19.91}}
\put(84.44,258.76){\vector(0,-1){20.14}}
\put(84.44,238.62){\vector(0,-1){19.68}}
\put(84.44,218.94){\vector(0,-1){20.14}}
\put(144.40,218.94){\vector(1,0){20.14}}
\linethickness{0.4pt}
\put(84.49,439.86){\circle*{4.30}}
\put(104.44,439.86){\circle*{4.30}}
\put(124.38,439.86){\circle*{4.30}}
\put(144.32,439.86){\circle*{4.30}}
\put(164.57,439.86){\circle*{4.30}}
\put(64.55,419.91){\circle*{4.30}}
\put(104.44,419.91){\circle*{4.30}}
\put(64.55,440.16){\circle*{4.30}}
\put(84.49,419.91){\circle*{4.30}}
\put(124.38,419.91){\circle*{4.30}}
\put(144.32,419.91){\circle*{4.30}}
\put(164.57,419.91){\circle*{4.30}}
\put(64.55,399.97){\circle*{4.30}}
\put(84.49,399.97){\circle*{4.30}}
\put(104.44,399.97){\circle*{4.30}}
\put(124.38,399.97){\circle*{4.30}}
\put(144.32,399.97){\circle*{4.30}}
\put(164.57,399.97){\circle*{4.30}}
\put(64.55,380.02){\circle*{4.30}}
\put(84.49,380.02){\circle*{4.30}}
\put(104.44,380.02){\circle*{4.30}}
\put(124.38,380.02){\circle*{4.30}}
\put(144.32,380.02){\circle*{4.30}}
\put(164.57,380.02){\circle*{4.30}}
\put(64.55,359.77){\circle*{4.30}}
\put(84.49,359.77){\circle*{4.30}}
\put(104.44,360.08){\circle*{4.30}}
\put(124.38,360.08){\circle*{4.30}}
\put(144.32,360.08){\circle*{4.30}}
\put(164.57,360.08){\circle*{4.30}}
\put(64.55,340.13){\circle*{4.30}}
\put(84.49,340.13){\circle*{4.30}}
\put(104.44,340.13){\circle*{4.30}}
\put(124.38,340.13){\circle*{4.30}}
\put(144.32,340.13){\circle*{4.30}}
\put(164.57,340.13){\circle*{4.30}}
\put(208.98,440.14){\circle*{4.30}}
\put(228.82,440.14){\circle*{4.30}}
\put(248.96,440.14){\circle*{4.30}}
\put(268.79,440.14){\circle*{4.30}}
\put(288.93,440.14){\circle*{4.30}}
\put(309.07,440.14){\circle*{4.30}}
\put(208.98,420.00){\circle*{4.30}}
\put(228.82,420.00){\circle*{4.30}}
\put(248.96,420.00){\circle*{4.30}}
\put(268.79,420.00){\circle*{4.30}}
\put(288.93,420.00){\circle*{4.30}}
\put(309.07,420.00){\circle*{4.30}}
\put(208.98,399.86){\circle*{4.30}}
\put(228.82,399.86){\circle*{4.30}}
\put(249.26,399.86){\circle*{4.30}}
\put(268.79,399.86){\circle*{4.30}}
\put(288.93,399.86){\circle*{4.30}}
\put(309.07,399.86){\circle*{4.30}}
\put(208.98,380.03){\circle*{4.30}}
\put(228.82,380.03){\circle*{4.30}}
\put(268.79,380.03){\circle*{4.30}}
\put(248.96,380.03){\circle*{4.30}}
\put(288.93,380.03){\circle*{4.30}}
\put(308.77,380.03){\circle*{4.30}}
\put(208.98,359.89){\circle*{4.30}}
\put(228.82,359.89){\circle*{4.30}}
\put(248.96,359.89){\circle*{4.30}}
\put(268.79,359.89){\circle*{4.30}}
\put(288.93,359.89){\circle*{4.30}}
\put(308.77,359.89){\circle*{4.30}}
\put(208.98,340.05){\circle*{4.30}}
\put(228.82,340.05){\circle*{4.30}}
\put(248.96,340.05){\circle*{4.30}}
\put(268.79,340.05){\circle*{4.30}}
\put(288.93,340.05){\circle*{4.30}}
\put(308.77,340.05){\circle*{4.30}}
\put(84.21,340.02){\line(0,-1){9.76}}
\put(164.27,340.02){\line(1,0){10.41}}
\put(104.39,340.02){\line(0,-1){9.11}}
\put(228.70,340.02){\line(0,-1){9.76}}
\put(248.88,340.02){\line(0,-1){9.76}}
\put(308.76,340.02){\line(1,0){9.76}}
\linethickness{0.5pt}
\bezier{56}(91.39,195.41)(89.11,189.12)(82.30,190.30)
\bezier{60}(82.30,190.30)(73.95,189.41)(72.50,195.40)
\bezier{64}(72.50,195.40)(71.09,203.99)(78.00,205.70)
\put(78.00,205.70){\vector(1,0){2.57}}
\linethickness{0.4pt}
\put(84.42,258.88){\circle*{4.30}}
\put(104.47,258.88){\circle*{4.30}}
\put(124.54,259.11){\circle*{4.31}}
\put(144.31,259.11){\circle*{4.30}}
\put(164.57,259.11){\circle*{4.30}}
\put(84.52,238.85){\circle*{4.30}}
\put(104.53,238.85){\circle*{4.30}}
\put(124.54,238.85){\circle*{4.30}}
\put(144.31,238.61){\circle*{4.31}}
\put(164.57,238.85){\circle*{4.30}}
\put(84.52,218.84){\circle*{4.30}}
\put(104.53,218.84){\circle*{4.30}}
\put(124.54,218.84){\circle*{4.30}}
\put(144.31,218.84){\circle*{4.30}}
\put(164.57,218.84){\circle*{4.30}}
\put(84.52,198.83){\circle*{4.30}}
\put(104.53,198.83){\circle*{4.30}}
\put(124.54,198.83){\circle*{4.30}}
\put(254.34,198.85){\vector(1,0){19.96}}
\linethickness{0.5pt}
\bezier{80}(264.03,194.44)(264.32,184.47)(254.34,185.00)
\bezier{76}(254.34,185.00)(245.83,184.47)(244.65,194.44)
\put(244.95,192.68){\vector(0,1){2.94}}
\linethickness{0.4pt}
\put(234.36,258.95){\circle*{4.30}}
\put(254.79,258.95){\circle*{4.30}}
\put(274.50,258.95){\circle*{4.30}}
\put(294.45,258.95){\circle*{4.30}}
\put(314.40,258.95){\circle*{4.31}}
\put(234.36,239.00){\circle*{4.31}}
\put(254.55,239.00){\circle*{4.31}}
\put(274.50,239.00){\circle*{4.31}}
\put(294.21,238.76){\circle*{4.30}}
\put(314.40,238.76){\circle*{4.30}}
\put(234.36,218.32){\circle*{4.30}}
\put(254.55,218.56){\circle*{4.30}}
\put(274.50,218.81){\circle*{4.30}}
\put(294.45,218.81){\circle*{4.30}}
\put(314.40,218.81){\circle*{4.30}}
\put(234.36,198.61){\circle*{4.30}}
\put(254.55,198.86){\circle*{4.30}}
\put(274.50,198.86){\circle*{4.30}}
\put(378.79,259.03){\circle*{4.30}}
\put(398.89,259.03){\circle*{4.30}}
\put(418.76,259.03){\circle*{4.30}}
\put(438.87,259.03){\circle*{4.30}}
\put(458.97,259.03){\circle*{4.30}}
\put(378.79,238.68){\circle*{4.30}}
\put(398.89,238.92){\circle*{4.30}}
\put(419.00,238.68){\circle*{4.30}}
\put(438.87,238.92){\circle*{4.30}}
\put(458.97,238.68){\circle*{4.30}}
\put(378.79,219.06){\circle*{4.30}}
\put(398.89,219.06){\circle*{4.30}}
\put(419.00,219.06){\circle*{4.30}}
\put(439.11,219.06){\circle*{4.30}}
\put(458.97,219.06){\circle*{4.30}}
\put(378.79,198.71){\circle*{4.30}}
\put(398.89,198.71){\circle*{4.30}}
\put(419.00,198.71){\circle*{4.30}}
\put(186.01,390.08){\makebox(0,0)[cc]{$\rightarrow$}}
\put(199.84,230.00){\makebox(0,0)[cc]{$\rightarrow$}}
\put(348.41,229.05){\makebox(0,0)[cc]{$\rightarrow$}}
\multiput(54.11,440.00)(2,0){60}{\makebox(0,0)[cc]{.}}
\multiput(54.44,420.00)(2,0){60}{\makebox(0,0)[cc]{.}}
\multiput(54.44,400.00)(2,0){60}{\makebox(0,0)[cc]{.}}
\multiput(54.44,380.00)(2,0){60}{\makebox(0,0)[cc]{.}}
\multiput(54.44,360.00)(2,0){60}{\makebox(0,0)[cc]{.}}
\multiput(54.44,340.00)(2,0){60}{\makebox(0,0)[cc]{.}}

\multiput(104.50,198.69)(0,2){30}{\makebox(0,0)[cc]{.}}
\multiput(124.33,238.86)(0,2){10}{\makebox(0,0)[cc]{.}}
\multiput(144.42,219.02)(0,2){20}{\makebox(0,0)[cc]{.}}
\multiput(84.420,219.27)(2,0){20}{\makebox(0,0)[cc]{.}}
\multiput(84.420,238.86)(2,0){10}{\makebox(0,0)[cc]{.}}
\multiput(144.42,238.86)(2,0){10}{\makebox(0,0)[cc]{.}}
\multiput(234.57,198.87)(2,0){10}{\makebox(0,0)[cc]{.}}
\multiput(234.57,218.82)(2,0){20}{\makebox(0,0)[cc]{.}}
\multiput(234.57,238.76)(2,0){10}{\makebox(0,0)[cc]{.}}
\multiput(254.51,198.87)(0,2){10}{\makebox(0,0)[cc]{.}}
\multiput(254.51,238.76)(0,2){10}{\makebox(0,0)[cc]{.}}
\multiput(274.45,238.76)(0,2){10}{\makebox(0,0)[cc]{.}}
\multiput(294.40,218.82)(0,2){20}{\makebox(0,0)[cc]{.}}
\multiput(378.97,218.93)(2,0){20}{\makebox(0,0)[cc]{.}}
\multiput(378.97,238.97)(2,0){20}{\makebox(0,0)[cc]{.}}
\multiput(438.85,238.97)(2,0){10}{\makebox(0,0)[cc]{.}}
\multiput(398.75,198.88)(0,2){10}{\makebox(0,0)[cc]{.}}
\multiput(398.75,238.97)(0,2){10}{\makebox(0,0)[cc]{.}}
\multiput(418.80,238.97)(0,2){10}{\makebox(0,0)[cc]{.}}

\multiput(64.44,450.00)(0,-2){60}{\makebox(0,0)[cc]{.}}
\multiput(84.44,450.00)(0,-2){60}{\makebox(0,0)[cc]{.}}
\multiput(104.44,450.00)(0,-2){60}{\makebox(0,0)[cc]{.}}
\multiput(124.44,450.00)(0,-2){60}{\makebox(0,0)[cc]{.}}
\multiput(144.44,450.00)(0,-2){60}{\makebox(0,0)[cc]{.}}
\multiput(164.44,450.00)(0,-2){60}{\makebox(0,0)[cc]{.}}

\multiput(314.34,238.76)(-2,0){10} {\makebox(0,0)[cc]{.}}
\multiput(438.85,218.93)(0,2){10}{\makebox(0,0)[cc]{.}}

\multiput(198.55,440.00)(2,0) {60}{\makebox(0,0)[cc]{.}}
\multiput(198.88,420.00)(2,0) {60}{\makebox(0,0)[cc]{.}}
\multiput(198.88,400.00)(2,0) {60}{\makebox(0,0)[cc]{.}}
\multiput(198.88,380.00)(2,0) {60}{\makebox(0,0)[cc]{.}}
\multiput(198.88,360.00)(2,0) {60}{\makebox(0,0)[cc]{.}}
\multiput(198.88,340.00)(2,0) {60}{\makebox(0,0)[cc]{.}}

\multiput(208.88,450.00)(0,-2){60}{\makebox(0,0)[cc]{.}}
\multiput(228.88,450.00)(0,-2){60}{\makebox(0,0)[cc]{.}}
\multiput(248.88,450.00)(0,-2){60}{\makebox(0,0)[cc]{.}}
\multiput(268.88,450.00)(0,-2){60}{\makebox(0,0)[cc]{.}}
\multiput(288.88,450.00)(0,-2){60}{\makebox(0,0)[cc]{.}}
\multiput(308.88,450.00)(0,-2){60}{\makebox(0,0)[cc]{.}}
\put(116.00,308.00){\makebox(0,0)[cc]{a}}
\put(259.00,308.00){\makebox(0,0)[cc]{b}}
\put(124.00,170.00){\makebox(0,0)[cc]{c}}
\put(274.00,171.00){\makebox(0,0)[cc]{d}}
\put(419.00,171.00){\makebox(0,0)[cc]{e}}
\put(90.63,366.92){\makebox(0,0)[cc]{$i_2$}}
\put(90.63,386.48){\makebox(0,0)[cc]{$i_3$}}
\put(110.80,366.92){\makebox(0,0)[cc]{$i_1$}}
\put(110.19,346.13){\makebox(0,0)[cc]{$i_0$}}
\put(234.91,386.48){\makebox(0,0)[cc]{$i_3$}}
\put(234.91,366.92){\makebox(0,0)[cc]{$i_2$}}
\put(255.69,366.92){\makebox(0,0)[cc]{$i_1$}}
\end{picture}
\caption{\ }
\end{figure*}

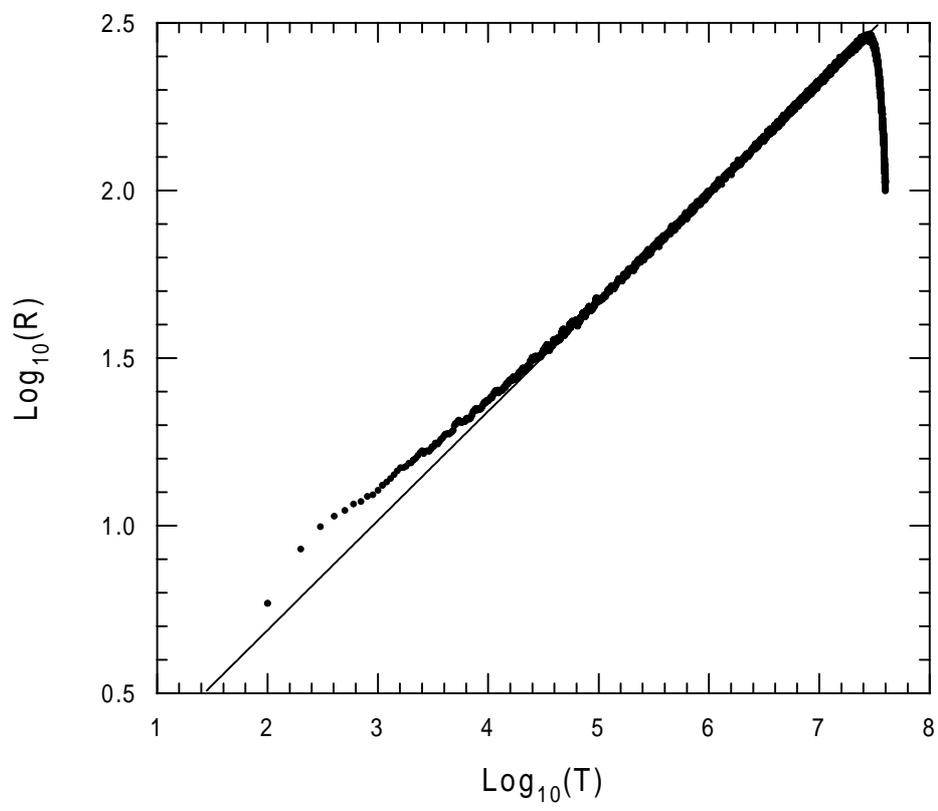
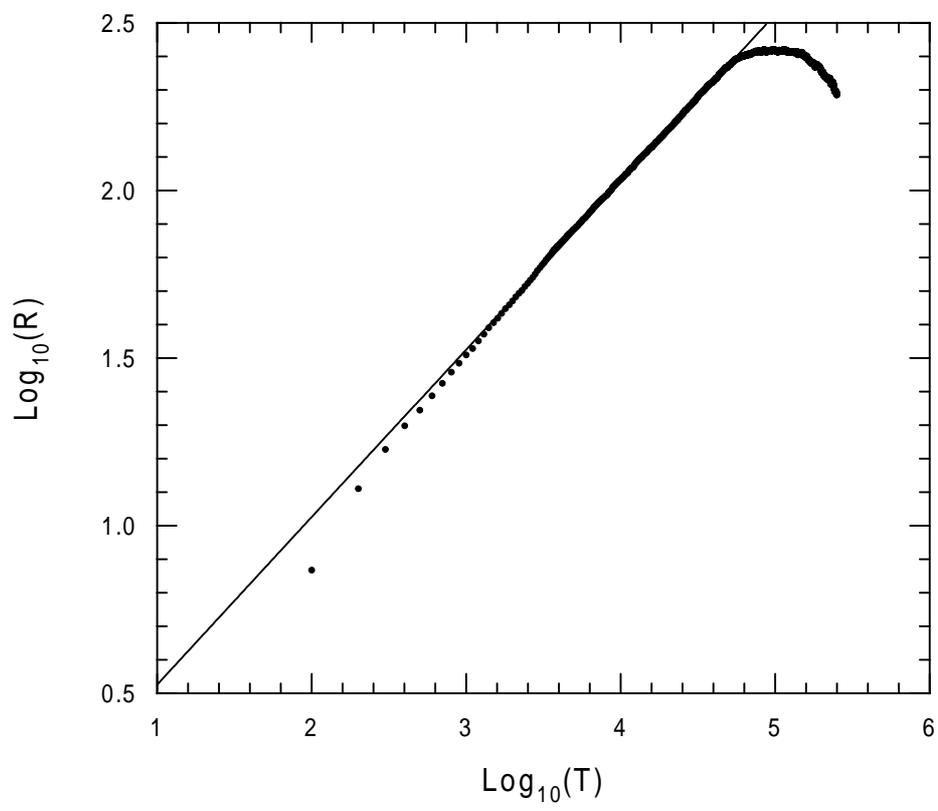
\end{document}